\documentclass[aps,prl,twocolumn,groupedaddress,showpacs]{revtex4}
\newcommand{\plb}[2]{{\em Phys.\ Lett.\/}                {\bf #1B}, #2 }
\newcommand{\plt}[2]{{\em Phys.\ Lett.\/}                {\bf  #1}, #2 }
\newcommand{\npb}[2]{{\em Nucl.\ Phys.\/}                {\bf B#1}, #2 }
\newcommand{\npp}[2]{{\em Nucl.\ Phys.\ Proc.\ Suppl.\/} {\bf #1C}, #2 }
\newcommand{\pan}[2]{{\em Phys.\ Atom.\ Nucl.\/}         {\bf  #1}, #2 }
\newcommand{\pr }[2]{{\em Phys.\ Rep.\/}                 {\bf  #1}, #2 }
\newcommand{\phr}[2]{{\em Phys.\ Rev.\/}                 {\bf  #1}, #2 }
\newcommand{\prt}[2]{{\em Phys.\ Rev.\/}                 {\bf D#1}, #2 }
\newcommand{\pru}[2]{{\em Phys.\ Rev.\ Lett.\/}          {\bf  #1}, #2 }
\newcommand{\rmf}[2]{{\em Rev.\ Mod.\ Phys.\/}           {\bf  #1}, #2 }

\newcommand{\con}[2]{                                    {\bf  #1}, #2 }
\newcommand{\etal}{{\em et al.}}
\newcommand{\ibid}{{\em ibid.\/}}
\newcommand{\ie}{{\em i.e.\ }}

\newcommand{\as}{\hat\alpha_s}

\newcommand{\be}{\begin{equation}}
\newcommand{\ee}{\end{equation}}
\newcommand{\ba}{\begin{eqnarray}}
\newcommand{\ea}{\end{eqnarray}}

\newcommand{\gsim}{\buildrel > \over {_\sim}}

\def\al2{\frac{\alpha^2}{\pi^2}}
\def\order#1{{\cal O}(#1)}

\begin{document}

\preprint{FT--2005--02}

\title{Upper Bound on the Hadronic Light-by-Light Contribution to the Muon 
$g - 2$}

\author{Jens Erler}
\author{Genaro Toledo S\'anchez}
\affiliation{Instituto de F\'\i sica, Universidad Nacional Aut\'onoma de 
             M\'exico, 01000 M\'exico D.F., M\'exico}

\date{\today}

\begin{abstract}
There are indications that hadronic loops in some electroweak observables are
almost saturated by parton level effects. Taking this as the hypothesis for 
this work, we propose a genuine parton level estimate of the hadronic 
light-by-light contribution to the anomalous magnetic moment of the muon, 
$a_\mu^{\rm LBL}({\rm had})$. Our quark mass definitions and values are 
motivated in detail, and the simplicity of our approach allows for
a transparent error estimate. For infinitely heavy quarks our treatment is 
exact, while for asymptotically small quark masses $a_\mu^{\rm LBL}({\rm had})$
is overestimated. Interpolating, this suggests to quote an upper bound. We 
obtain $a_\mu^{\rm LBL}({\rm had}) < 1.59\times 10^{-9}$ (95\%~CL).
\end{abstract}


\maketitle


The E--821 Collaboration at BNL~\cite{Bennett:2004pv} has measured 
the anomalous magnetic moment of the muon, $a_\mu \equiv (g_\mu - 2)/2$, with
an uncertainty of $\Delta a_\mu = \pm 0.63\times 10^{-9}$. This would provide
sensitivity to new physics scales, $\Lambda$, as large as 
$\Lambda \sim m_\mu/\sqrt{\Delta a_\mu} = 4.2$~TeV, where $m_\mu$ is the mass
of the muon~\cite{Czarnecki:2001pv}. A new experiment~\cite{Hertzog:2005mx} at 
BNL is approved aiming at even greater ($\pm 0.2$~ppm) precision and implying 
a reach up to 7~TeV. Unfortunately, the interpretation of $a_\mu$ is 
compromised by large theoretical uncertainties introduced by hadronic effects 
diluting the new physics sensitivity. These mainly arise from two-loop vacuum 
polarization (VP) effects, $a_\mu^{\rm VP}({\rm 2,had})$, and from 
the three-loop contribution of light-by-light type, 
$a_\mu^{\rm LBL}({\rm had})$. The calculations of $a_\mu^{\rm LBL}({\rm had})$ 
based on chiral perturbation theory~\cite{Knecht:2001qg,Ramsey-Musolf:2002cy} 
are the only ones to date solidly based on a systematic expansion. However, 
the estimated uncertainty, $\gsim\pm 10^{-9}$, is significantly 
larger~\cite{Ramsey-Musolf:2002cy} than the measurement error. There is 
a variety of model estimates, all supplementing the dominant $\pi^0$-exchange 
contribution with other resonance exchanges and $\pi^\pm$-loops. Current 
analyzes~\cite{Knecht:2001qf,Melnikov:2003xd} agree reasonably well on 
the magnitude (where residual differences are largely understood) and the sign 
of $a_\mu^{\rm LBL}({\rm had})$. They have reached a high level of 
sophistication, but the error estimates remain rough guesses. There is also 
an estimate~\cite{Dorokhov:2005ff} based on the instanton liquid model with 
a small ($ < 10\%$) quoted error. 

In this work we offer a complementary way to estimate 
$a_\mu^{\rm LBL}({\rm had})$, with no need to commit to the dominance of any 
particular type of contribution. It is a na\"ive parton level estimate which is
solid in the heavy quark limit where it matches onto perturbative QCD, 
but overestimates the light-by-light contribution in the chiral limit. 
Therefore, our approach naturally implies an {\em upper bound\/} for 
$a_\mu^{\rm LBL}({\rm had})$, which is very useful in view of the $a_\mu$ 
measurement lying above the Standard Model prediction. It can also be applied 
to $a_\mu^{\rm VP}({\rm had})$, which serves as a reference case and allows for
a {\em transparent\/} error estimate. 

Attempting to obtain $a_\mu^{\rm LBL}({\rm had})$ directly at the parton level
seems hopeless at first since perturbative QCD cannot be applied except for 
the heavy $c$ and $b$ quarks. Moreover, in the chiral limit, in which the bulk 
of the effect arises from light, charged, and essentially point-like 
pseudo-Goldstone bosons, one has $a_\mu^{\rm LBL}({\rm had}) < 0$ from scalar 
QED~\cite{Kuhn:2003pu}, while leptons and heavy quarks contribute 
positively~\cite{Laporta:1992pa}. However, as shown in 
Ref.~\cite{Melnikov:2003xd}, the $\pi^\pm$-loops receive large chiral symmetry 
breaking corrections, rendering the net $\pi^\pm$-contribution almost 
negligible. Thus, at least as far as $a_\mu^{\rm LBL}({\rm had})$ is concerned,
we are dealing with a kinematic regime that conceivably admits an alternative 
description in terms of quark in place of chiral degrees of freedom. Indeed, 
an estimate~\cite{Melnikov:2003xd} of the typical momenta, $\mu$, in 
the $\pi^\pm$-loops yields $\mu\sim 4.25\ m_\pi\approx 0.6$~GeV, \ie large 
enough to feel individual partons even within pions. These observations revive 
the question whether hadronic light-by-light diagrams are best described by 
hadrons or by partons or a mix. We are aware of three parton based attempts in 
the literature. An early approach~\cite{Calmet:1976pu} using exclusively 
constituent quarks~\cite{DeRujula:1975ge} resulted in a numerical estimate,
albeit with the wrong sign. Ref.~\cite{DeTroconiz:2001wt} opted for a mixed 
treatment in which constituent quark masses, $m_{u,d} = 0.33$~GeV, 
$m_s = 0.5$~GeV, and $m_c = 1.87$~GeV (with assumed errors of $\pm 10\%$), were
employed in the analytical result of Ref.~\cite{Laporta:1992pa} and produced 
the correct sign. The adopted definition, values, and uncertainty were not 
further justified, but in any case the constituent quarks were assumed to yield
only the high momentum contribution which was supplemented with 
$\pi^0$-exchange graphs as described by the chiral constituent quark 
model~\cite{Manohar:1983md}. Ref.~\cite{Pivovarov:2001mw} is very close to both
our strategy and philosophy and will be commented on below 
relation~(\ref{limit}).

Indeed, a main obstacle for any model involving quark degrees of freedom at 
hadronic scales is to find the appropriate quark mass {\em definition\/}. In 
contrast to pole masses, $\overline{\rm MS}$ masses, $\hat{m}_q(\mu)$ (where 
$\mu$ is the renormalization scale), tend to produce small perturbation 
coefficients (at least for sufficiently inclusive quantities). We will 
therefore avoid pole masses for heavy quarks and similarly constituent masses 
for light quarks. The scale choice is implied by the requirement of avoiding 
spurious logarithms, $\ln\mu^2/\hat{m}^2_q(\mu)$ or $\ln\mu^2/m^2_\mu$. They 
can be nullified (resummed) by choosing $\mu = \hat{m}_q(\mu)$ for 
$\hat{m}_q(\hat{m}_q) > m_\mu$ and $\mu = m_\mu$ for 
$\hat{m}_q(\hat{m}_q) < m_\mu$, \ie the higher of the two masses sets 
the scale. There are also genuine logarithms of the form 
$\ln\hat{m}^2_q(\mu)/m^2_\mu$ which cannot be eliminated.

But for the light ($u,d,s$) quark contribution to $g_\mu-2$, the values of 
$\hat{m}_q(\hat{m}_q)$ or $\hat{m}_q(m_\mu)$ are unknown because $\as$ and 
consequently the anomalous quark dimensions formally diverge. In a similar 
context, however, it was possible to proceed phenomenologically. Namely, 
the hadronic contribution to the renormalization group evolution (RGE) of 
the QED coupling (the running), $\hat\alpha(\mu)$, can be 
mimicked~\cite{Marciano:1980pb} by effective quark masses. These have been 
defined more precisely in Ref.~\cite{Erler:2004in} as {\em threshold quark
masses\/}, $\bar{m}_q$, and correspond to the scale at which the theory is
changed to include or exclude the quark $q$ and where the one-loop QCD $\beta$ 
function coefficient changes correspondingly. Since this is {\em not\/} 
a perturbative description, higher order QCD effects are assumed to be absorbed
in the threshold masses, leading to~\cite{Erler:2004in},
\begin{eqnarray}
\label{udmass}
    \bar{m}_u =  \bar{m}_d &=& 176 \pm 9 \mbox{ MeV}, \\
\label{smass}
    \bar{m}_s &=& 305^{-65}_{+82} \mbox{ MeV}, \\
\label{cmass}
    \bar{m}_c &=& 1.18 \pm 0.04 \mbox{ GeV}, \\
\label{bmass}
    \bar{m}_b &=& 4.00 \pm 0.03 \mbox{ GeV}, 
\end{eqnarray}
where isospin symmetry, $\hat{m}_u(\hat{m}_u) = \hat{m}_d(\hat{m}_d)$, was 
imposed. The uncertainties in the first (last) two lines have a correlation of
$-100\%$ (+29\%). 
The values~(\ref{udmass}--\ref{bmass}) are implied by 
the considerations~\cite{Erler:2004in} in the next paragraph.

Heavy quark masses are qualitatively given by one half of the mass, $M_{1S}$,
of the corresponding $1S$ quarkonium resonance. If the quark is heavy enough 
that QCD perturbation theory can be applied, this statement can be made 
quantitative resulting in well studied QCD sum 
rules~\cite{Novikov:1977dq,Erler:2002bu}, which are dominated by the 
$1S$-quarkonium resonance and supplemented by higher vector resonances and 
a continuum contribution. They yield expressions for $\hat{m}_q(\hat{m}_q)$ 
with small QCD corrections which are known~\cite{Chetyrkin:1996cf} and included
up to order $\alpha_s^2$. The threshold masses in 
Eqs.~(\ref{cmass}--\ref{bmass}) can then be obtained in terms of these using 
Eq.~(32) in Ref.~\cite{Erler:2004in}.
Defining the parameter, $\xi_q = 2 \bar{m}_q/M_{1S}$~\cite{Erler:2004in}, we 
have that asymptotically $\xi_q \to 1$ for $\bar{m}_q \to \infty$ and 
$\xi_q\to 0$ for $\bar{m}_q\to 0$. Thus, for a heavy quark $\xi_q\sim 1$, while
for a light quark $\xi_q \ll 1$. Also, we expect 
$\bar{m}_1 < \bar{m}_2 \Longrightarrow \xi_1 < \xi_2$, which can be checked 
explicitly by comparing $\xi_b = 0.845 \pm 0.006$ with $\xi_c = 0.76 \pm 0.03$.
This implies the upper bound $\xi_s < \xi_c$. A lower bound, $\xi_s > 0.47$, 
can be obtained by considering the $SU(3)$ limit, $\xi_u = \xi_d = \xi_s$. 
Since the total contribution to the running of $\alpha$ due to $u$, $d$, and 
$s$ quarks is known experimentally~\footnote{The corresponding experimental 
error induces an additional uncertainty into $\bar{m}_u$, $\bar{m}_d$, and 
$\bar{m}_s$. This is an order of magnitude below the one in 
Eqs.~(\ref{udmass}--\ref{smass}) and neglected.}, this (together with isospin 
symmetry) implies Eq.~(\ref{smass}). Assuming in addition the suppression of 
flavor singlet contributions (the OZI rule~\footnote{This assumption is not 
necessary to derive the $SU(3)$ limit, because in this case the flavor singlet
contribution vanishes by virtue of $Q_u + Q_d + Q_s = 0$. OZI rule violations 
could still affect the result~(\ref{udmass}) but on general grounds we expect 
them to be tiny~\cite{Erler:2004in}.}) then implies Eq.~(\ref{udmass}) and 
explains the complete anti-correlation with Eq.~(\ref{smass}). Isospin 
violations will, of course, induce a small mass difference, 
$\bar{m}_d - \bar{m}_u$.

In Ref.~\cite{Erler:2004in} the threshold masses~(\ref{udmass}--\ref{bmass})
were used for the RGE of the weak mixing angle, $\sin^2\hat\theta(\mu)$, which
is rigorously justified as the same vector current correlator convoluted by
the same weight function enters in both, $\hat\alpha(\mu)$ and 
$\sin^2\hat\theta(\mu)$. What we are proposing here, is that 
$a_\mu^{\rm LBL}({\rm had})$ can be {\em modeled\/} by treating quarks like
massive leptons and using the values~(\ref{udmass}--\ref{bmass}). Notice, that
these values are systematically lower than typical constituent
masses~\cite{DeRujula:1975ge}, which is reasonable since we do not add 
an independent contribution from pion degrees of freedom.

It is {\em a priori\/} not obvious that this procedure and these values for 
the quark masses (in fact, any set of values) will generate the correct 
$a_\mu^{\rm LBL}({\rm had})$, but this represents the only {\em ad hoc\/} model
assumption that we need to make. One justification is the observation mentioned
before, that the relevant kinematic region conceivably admits a description in 
terms of quark degrees of freedom, which is reinforced by the lower bounds, 
$\xi_s > 0.47$ and $\xi_{u,d} > 0.42$, showing that the physical light quark 
masses are midway between the perturbative ($\xi\to 1$) and chiral ($\xi\to 0$)
limits. Thus, at a {\em qualitative\/} level, both $a_\mu^{\rm LBL}({\rm had})$
and the RGE of $\hat\alpha(\mu)$ appear to be describable by moderately heavy 
quark degrees of freedom. 

Although our treatment clearly improves with growing $\bar{m}_q$ and becomes 
exact for $\bar{m}_q \to \infty$, it is important to establish 
{\em quantitative\/} tests and to apply our approach to physical quantities 
for which the answer is known with sufficient precision. For this we note that 
$a_\mu^{\rm VP}({\rm 2,had})$ is described by the same vector current 
correlator as the RGE of $\hat\alpha(\mu)$ but is convoluted with a different 
weight function, $K^{(1)}(s)$~\cite{Brodsky:1967sr}, so that just as in 
the case of $a_\mu^{\rm LBL}({\rm had})$ it is {\em a priori\/} not obvious 
that our approach and quark mass values will generate the correct answer,
\be
\label{dispresult}
   a_\mu^{\rm VP}({\rm 2,had})\times 10^{9} = 70.30 \pm 0.43 \pm 0.88 
                                            = 70.3  \pm 1.0
\ee
To arrive at Eq.~(\ref{dispresult}) we averaged the results based on $e^+ e^-$ 
annihilation data and $\tau$ decay data which are both taken from 
Ref.~\cite{Davier:2005xq}. The second uncertainty accounts for the discrepancy 
between the two data sets. It is chosen to one half of the difference in 
central values and added in quadrature to the first (experimental) one. 
We compare this result with the analytical 
expression~\cite{Elend:1966,Samuel:1990qf},
\be
\label{amuvpana}
   a_\mu^{\rm VP}({\rm 2,had}) = {\alpha^2\over\pi^2} N_C \sum\limits_q Q_q^2 
   \sum\limits_{n=0}^{\infty} \left( {m_\mu\over\bar{m}_q} \right)^{2n+2}\times
\ee
\vspace{-12pt}
$$
   \left[ {n \ln ({m_\mu^2/\bar{m}_q^2}) \over (n + 3)(2 n + 3)(2 n + 5)} - 
   {8 n^3 + 28 n^2 - 45\over [(n + 3)(2 n + 3)(2 n + 5)]^2} \right],
$$
yielding, 
\be
\label{amuvp}
   a_\mu^{\rm VP}({\rm 2,had}) = 65.1^{-2.6}_{+4.2} \mp 0.1\times 10^{-9},
\ee
The first and second uncertainty are induced, respectively, from 
Eqs.~(\ref{udmass}--\ref{smass}) and Eqs.~(\ref{cmass}--\ref{bmass}). Effects 
due to $\bar{m}_d \neq \bar{m}_u$ cancel to first order in 
$a_\mu^{\rm VP}({\rm 2,had})$ (but not in $a_\mu^{\rm LBL}({\rm had})$ below) 
and can be neglected. Notice, that the estimate~(\ref{amuvp}) is in reasonable 
agreement with Eq.~(\ref{dispresult}) and that our central value is 
{\em below\/} it. This is consistent with the chiral limit, $m_\pi\to 0$ with 
$m_\mu/m_\pi$ fixed, underestimating~\footnote{The alternative chiral limit, 
$m_\pi\to 0$ with $m_\mu$ fixed, is reproduced correctly, because then 
the leading effects in both $a_\mu^{\rm VP}({\rm 2,had})$ and the RGE of 
$\hat\alpha(\mu)$ are logarithmic.} $a_\mu^{\rm VP}({\rm 2,had})$. Indeed, in 
this limit,
$$
   m_\pi\to 0 \Longrightarrow {\bar{m}_q\over\mu} \to 
   \left( {m_\pi\over\mu}\right)^{\sqrt{1\over 4 N_C (Q_u^2+Q_d^2)}} \approx
   \left( {m_\pi\over\mu}\right)^{0.39},
$$
(where $\mu$ is a fixed reference scale) so that $m_\mu/\bar{m}_q\to 0$. 
Combined with $K(s) > 0$ this implies a systematic {\em underestimate\/}. 
The fact that the estimate~(\ref{amuvp}) reproduces the experimental result 
within about $10\%$ provides phenomenological support for our approach and its 
use of quark degrees of freedom. We can also directly compare the contribution 
from $c$ quarks, $a_\mu^{\rm VP}(2,c) = 1.27\times 10^{-9}$ from 
Eqs.~(\ref{cmass}) and (\ref{amuvpana}), with the analytical result up to order
$\alpha_s^2$ originally obtained in the second Ref.~\cite{Chetyrkin:1996cf} in 
the form given in Eq.~(4) of Ref.~\cite{Erler:2000nx}. Using 
$\hat{m}_c(\hat{m}_c) = 1.29$~GeV and $\as(M_Z) = 0.1216\pm 0.0017$ 
(corresponding to $\as(\hat{m}_c) = 0.432$ and $\as(\bar{m}_c) = 0.473$), 
the latter yields $a_\mu^{\rm VP}(2,c) = 1.46\times 10^{-9}$. Thus, here we 
underestimate the known result by 13\%, which is traceable to different 
$\alpha_s$ dependences in the RGE of $\hat\alpha(\mu)$ and the leading 
$\order{m_\mu^2/\hat{m}_c^2}$ term in $a_\mu^{\rm VP}({\rm 2,had})$. QCD 
corrections to the $\order{m_\mu^4/\hat{m}_q^4}$ term which are significant 
only for lighter quarks tend to have a compensating effect, offering 
a perturbative rationale why our approach seems to work better for light quarks
than for charm. 

The success of the test in the previous paragraph may be coincidental. 
Therefore, we performed three similar tests on various three-loop VP effects, 
$a_\mu^{\rm VP}({\rm 3i,had})$, each with a different kernel function,
$K^{(2i)}(s)$~\cite{Barbieri:1974nc,Calmet:1976kd}. $i=a$ contains an 
additional photonic correction or muon-loop relative to 
$a_\mu^{\rm VP}({\rm 2,had})$, while $i=b$ $(c)$ corresponds to an additional 
electron (hadron) loop insertion. Our method (in parentheses) reproduces 
the central values of Ref.~\cite{Krause:1996rf}, 
\begin{equation}
\begin{array}{lrr}
 a_\mu^{\rm VP}({\rm 3a,had}) =& -2.11\times 10^{-9} & (-1.82\times 10^{-9}),\\
 a_\mu^{\rm VP}({\rm 3b,had}) =& +1.07\times 10^{-9} & (+0.99\times 10^{-9}),\\
 a_\mu^{\rm VP}({\rm 3c,had}) =& +2.7\times 10^{-11} & (+2.8\times 10^{-11}),
\end{array}
\end{equation}
within 14\%, 7\%, and 4\%, respectively. Note, that these represent genuine 
``hit or miss'' tests involving no adjustable parameters, and all four passed
at the 10\% level.

We finally apply our approach to $a_\mu^{\rm LBL}({\rm had})$. Our master 
formula is the one for heavy leptons~\cite{Laporta:1992pa},
$$
   a_\mu^{\rm LBL}({\rm had}) = {\alpha^3\over\pi^3} N_C \sum\limits_q Q_q^4
   \sum\limits_{n=1}^\infty {m_\mu^{2n}\over\bar{m}_q^{2n}}
   \sum\limits_{k=0}^2 C_{n,k} \ln^k{m_\mu^2\over\bar{m}_q^2},
$$
where $C_{1,0} = 3/2 \zeta_3 - 19/16$, $C_{1,1} = C_{1,2} = 0$, and where 
the other $C_{n,k}$ up to $n = 5$ can be found in Eq.~(5) of 
Ref.~\cite{Kuhn:2003pu}. The central values in Eqs.~(\ref{udmass}--\ref{bmass})
then imply $a_\mu^{\rm LBL}({\rm had}) = 1.36 \pm 0.13\times 10^{-9}$, where 
the uncertainty is the model error, determined by the 9.5\% difference between 
the upper end of the range~(\ref{dispresult}) from the central value in 
Eq.~(\ref{amuvp}). Similarly, the upper and lower error in 
Eqs.~(\ref{udmass}--\ref{smass}) and (\ref{amuvp}) correspond to 
$a_\mu^{\rm LBL}({\rm had}) = 1.27 \pm 0.18\times 10^{-9}$ and
$a_\mu^{\rm LBL}({\rm had}) = 1.47 \pm 0.04\times 10^{-9}$, \ie model errors of
14.1\% and 2.9\%, respectively. As expected, $b$ quark effects are negligible, 
but the contribution from $c$ quarks (usually ignored) turns out to be 
$0.04\times 10^{-9}$. Although small it might serve as an interesting testing 
tool of other model calculations of $a_\mu^{\rm LBL}({\rm had})$ which should 
be consistent with what we have obtained, since our method is solid (up to 
effects of $\order{\alpha_s}$) for heavy quarks. To account for isospin 
breaking we shift $\bar{m}_d$ by +2~MeV and $\bar{m}_u$ by $-1$~MeV, increasing
all results by $0.01\times 10^{-9}$ to
\be
\label{amulbl2}
   a_\mu^{\rm LBL}({\rm had}) = 1.37^{-0.27}_{+0.15}\times 10^{-9},
\ee
where the error covers the three ranges above and is constructed to include 
the intrinsic model uncertainty. As in any hadronic model there would be 
additional model uncertainty if the comparison to other cases (here 
$a_\mu^{\rm VP}({\rm had})$) was regarded as pure coincidence. The result, 
$1.36 \pm 0.25\times 10^{-9}$, of Ref.~\cite{Melnikov:2003xd} is the first 
to take short-distance QCD constraints explicitly into account. By 
construction, this is a feature of our approach, as well, so it is gratifying 
that the estimate~(\ref{amulbl2}) turns out to agree perfectly. Our central 
value is {\em higher\/} than previous ones~\cite{Knecht:2001qf}, which is 
expected given that in the chiral limit, $m_\pi\to 0$ with $m_\mu/m_\pi$ fixed 
(and presumably also with $m_\mu$ fixed, but the scalar QED result in this case
is not known), we {\em overestimate\/} $a_\mu^{\rm LBL}({\rm had})$ since 
$\pi^\pm$-loops contribute negatively. 

Now we repeat the above analysis with the model errors multiplied by 1.645 to
shift from uncertainties estimated in a ``$1\sigma$ spirit'' to 90\% ranges,
and find $a_\mu^{\rm LBL}({\rm had}) = 1.37 \pm 0.21\times 10^{-9}$,
$a_\mu^{\rm LBL}({\rm had}) = 1.28 \pm 0.30\times 10^{-9}$, and
$a_\mu^{\rm LBL}({\rm had}) = 1.48 \pm 0.07\times 10^{-9}$, respectively.
The upper errors have converged, and since as mentioned our approach tends 
to overestimate $a_\mu^{\rm LBL}({\rm had})$, we quote as our final result 
the 95\%~CL upper bound,
\be
\label{limit}
   a_\mu^{\rm LBL}({\rm had}) < 1.59\times 10^{-9}.
\ee

In a similar paper~\cite{Pivovarov:2001mw}, a 1-parameter fit to 
$a_\mu^{\rm VP}({\rm 2,had})$ resulted in 
$m_u = m_d = m_s - 180\mbox{ MeV} = 166 \pm 1$~MeV. These values are within 
errors consistent with Eqs.~(\ref{udmass}--\ref{smass}), reflecting our 
statement that $a_\mu^{\rm VP}({\rm 2,had})$ and $\hat\alpha(\mu)$ can be 
approximated using the same masses. We benefited from a deeper 
$m_\mu^2/m_q^2$-expansion of $a_\mu^{\rm LBL}$~\cite{Kuhn:2003pu}, lowering 
the light quark contribution relative to~\cite{Pivovarov:2001mw} by 
$\sim 6$\%. The most important new aspect here is that the recent scalar QED 
calculation~\cite{Kuhn:2003pu} allows us to interpolate between the chiral and 
heavy quark limits. If one can prove (conceivably by lattice simulations) 
the monotony of the product $m_q^2 a_\mu^{\rm LBL}({\rm had})$ as a function of
$m_q$, our bound~(\ref{limit}) would become rigorous, but this remains 
a loophole for now.

We stress that it is worthwhile to compute the $\alpha_s$ correction to
the heavy quark light-by-light contribution to $a_\mu$ to check explicitly its
magnitude and more importantly its sign, to compare it to the RGE of
$\hat\alpha(\mu)$, and to use it as a refinement of our method. 
This contribution may be called
$A^{(6,2)}_2(m_\mu/\hat{m}_q)$ and corresponds to a subclass of 60 of the 
469 Feynman diagrams contributing at $\order{\alpha^4}$ to what is known as 
$A^{(8)}_2(m_\mu/m_\tau)$ with $m_\tau$ replaced by $\hat{m}_q(\hat{m}_q)$ and
one prefactor of $\alpha$ by $\hat\alpha_s(\hat{m}_q)$. This is possible with 
state-of-the-art computer codes~\cite{Kinoshita:2004wi}. 

We have developed (quark mass definition, scale settings, quark mass values, 
and uncertainties) what we believe are necessary ingredients for a parton model
estimate of $a_\mu^{\rm LBL}({\rm had})$ and similar quantities. The model has
four key benefits: (i) it is simple and transparent, (ii) it can be tested, as
we did successfully using $a_\mu^{\rm VP}({\rm had})$, (iii) short-distance 
constraints are taken into account by construction, and (iv) it naturally 
provides an upper bound on $a_\mu^{\rm LBL}({\rm had})$ because, as we have 
argued, our method tends to yield an overestimate and also because the 95\% 
upper error is virtually insensitive to our input masses.

We would like to compare our approach with celebrated models of low-energy QCD,
such as the vector meson dominance model~\cite{Nambu:1962}, the $1/N_C$ 
expansion~\cite{'tHooft:1974hx}, the chiral constituent quark 
model~\cite{Manohar:1983md}, or the extended~\cite{Bijnens:1992uz} 
Nambu-Jona-Lasinio~\cite{Nambu:1961tp} model. It is not obvious 
{\em a priori\/} that approximations like $N_C = 3 \gg 1$~\cite{'tHooft:1974hx}
or $\Lambda_{\rm QCD}\ll M_{\chi SB}$ (the spontaneous chiral symmetry breaking
scale)~\cite{Manohar:1983md}, or the dominance of certain 
resonances~\cite{Nambu:1962} or operators~\cite{Bijnens:1992uz} --- while 
plausible --- would generate useful results. But experience showed that they 
do. The assumption made in this work is that for the type of observables 
considered, parton level effects dominate and almost saturate hadronic loops. 
It is in line with the observations of Ref.~\cite{Melnikov:2003xd}, and we 
found further evidence for it in both, $a_\mu^{\rm VP}({\rm had})$ and 
$a_\mu^{\rm LBL}({\rm had})$, which passed quantitative tests with better than 
expected phenomenological success. This may be interpreted as pure coincidence 
and, clearly, it is important to submit our method to other case studies in 
the future. More optimistically, however, our results offer an alternative view
on hadronic loops and may lead to a rigorous bound on 
$a_\mu^{\rm LBL}({\rm had})$.

\begin{acknowledgments}
It is a pleasure to thank Michael Ramsey-Musolf and Lee Roberts for helpful
comments on the manuscript and Lorenzo D\'\i az-Cruz and Peter Zerwas for 
fruitful discussions. JE greatly acknowledges the hospitality of Caltech, DESY 
(Hamburg), and U.\ Penn. This work was supported by CONACyT (M\'exico) contract
42026--F and by DGAPA--UNAM contract PAPIIT IN112902.
\end{acknowledgments}

\end{document}